\documentclass[aps,pre,epsfigure,twocolumn,superscriptaddress,nofootinbib]{revtex4-1}

\usepackage{amssymb, theorem, xspace}
\usepackage{float}
\usepackage{dcolumn}
\usepackage{slashed}
\usepackage{tikz}
\usetikzlibrary{quantikz}
\usepackage[colorlinks=true,linkcolor=blue,urlcolor=blue,citecolor=blue,pdfusetitle]{hyperref}
\usepackage[utf8]{inputenc}
\usepackage[english]{babel}
\usepackage{subfigure}
\usepackage{blindtext}
\usepackage{amsfonts}
\usepackage{bbm}
\usepackage{enumerate}
\usepackage{color}

\begin{document}
\title{Symplectic Representation of the Ginzburg-Landau Theory}

\author{E.A. Reis$^{1,2}$, G.X.A. Petronilo$^{1,3}$, R.G.G. Amorim$^{1,4,8}$\\
H. Belich$^{5}$,  F.C. Khanna$^{6,7}$, A.E. Santana}
\address{
International Center of Physics, Instituto de F\'isica, \\ Universidade de Bras\'ilia, 70.910-900, Brasilia, DF, Brazil.\\
$^{2}$Centro das Ci\^encias Exatas e das Tecnologias, \\ Universidade Federal do Oeste da Bahia, 47.808-021, Barreiras, BA, Brazil. \\
$^{3}$Latin American Quantum Computing Center, 
41650-010, Salvador, BA, Brazil.\\
 $^{4}$Department of Physics, Gamma Faculty, \\ University of Bras\'ilia, 72.444-240, Bras\'ilia, DF, Brazil.\\
$^{5}$Centro de Ci\^encias Exatas, Universidade Federal do Esp\'irito Santo, \\ 29.060-900, Vit\'oria, ES, Brazil.\\
$^{6}$Physics Department, Theoretical Physics Institute, \\ University of
Alberta, Edmonton, AB, T6G 2J1, Canada.\\
$^{7}$TRIUMF, 4004, Westbrook mall,  V6T 2A3, Vancouver, BC,
Canada.\\
$^{8}$Canadian Quantum Research Center,\\
204-3002 32 Ave Vernon, BC V1T 2L7  Canada
}

%\date{\today}

\begin{abstract}
In this work, the Ginzburg-Landau theory is represented on a symplectic manifold with a phase space content. The order parameter is defined by a quasi-probability amplitude, which gives rise to a quasi-probability distribution function, i.e., a Wigner-type function. The starting point is the thermal group representation of Euclidean symmetries and gauge symmetry. Well-known basic results on the behavior of a superconductor are re-derived, providing the consistency of representation. The critical superconducting current density is determined and its usual behavior is inferred. The negativety factor associated with the quasi-distribution function is analyzed, providing information about the  non-classicality nature of the superconductor state in the region closest to the edge of the superconducting material.
\\
\textbf{keywords}: Thermo-Euclidian group; Ginzburg-Landau theory; Phase space.
\end{abstract}

\maketitle
\section{Introduction}
One of the most fascinating phenomena in physics, superconductivity, was first experimentally observed by Onnes~\cite{Onnes}. Since then phenomenological models have been an important tool to provide  explanations of the superconducting matter state. Initially, this was due to the impossibility to consider the problem from a classical perspective only.  London brothers~\cite{London} arrived at penetration length, based on the work of Meissner and Ochsenfeld~\cite{Meissner} and by using a phenomenology centered  entirely on classical elements.  Ginzburg and Landau~\cite{GinzburgLandau1950} proposed a phenomenological formalism considering characteristics of the fluctuation associated with the order parameter of the superconducting state. The model was a generalization of  the  phase-transition Landau theory. The Ginzburg-Landau  model describes in particular the Meissner effect of the magnetic field expulsion, the penetration length and also the so-called coherence length.

Due to the success, other studies have been carried out over the decades  to improve and clarify general concepts about superconducting systems. These studies consider, in particular, the nonlinear nature of the Ginzburg-Landau free energy. Then it should be of interest to examine the consistency of  critical phenomena from the starting point of phase space with the perspective to explore elements of non-classicality and chaoticity. In this context, it is remarkable the analysis  based on  symplectic  topological manifolds, giving rise, as an example, to Poincar\'e maps. However, it is far from being fully explored~\cite{Contoyiannis}.

It is  important to emphasize at this point that fluctuations in  the Ginzburg-Landau model are introduced in parallel to the case of quantum mechanics. This aspect points to the  possibility of representing the order parameter in association with a quasi-probability distribution; that is, a Wigner-like function. Then the singular importance of the Wigner function for the analysis of quantum chaoticity, dissipation and non classicality~\cite{Korsch,Hutchinson,Greenbaum,Ademir2009} would provide an expanded view about the behavior of the order parameter at critical points. 
This is a central goal in the present work. We start by formulating the problem by using a symplectic Hilber space and representations of symmetry groups.
In other words, since the Ginzburg-Landau order parameter is a field  and   the free energy density (equivalent to a Lagrangian density in field theory) is invariant by Euclidian-group transformations,  it is possible to derive symplectic representations for this theory.

Regarding the use of symplectic structures  to study non-relativistic and relativistic field theories, several approaches have been proposed~\cite{1TorresVega1990,1Dito1992,1Gosson2005,1chru1,1Gosson2008,1Gosson2009}, in particular starting from symmetry groups~\cite{Ronni2007,Ronni2019,Hara2022,Renato2022}. The starting point is the Lie algebra, which leads to  equations of  motion describing physical systems in phase space.  In this case, a central element is the  Wigner function $f_{W}(q,p)$, where $q$ and $p$ stand for a position point and the canonical momentum, respectively, both defined in the Euclidian metric space.~\cite{Wigner}.

The Wigner function was proposed  as an attempt to improve quantum corrections to the classical kinetic theory~\cite{Wigner}. Although $f_{W}(q,p)$ is not a probability distribution,   observable averages in quantum mechanics are carried out in a similar way  to those used in classical physics~\cite{1schroeck,1ferrya,1curtrbook,1moya,1kimzacha,1kimnoz,1Wignerconc}. In the Wigner approach, each operator $A$, defined in (Euclidian) Hilbert space $\mathcal{H}_E$, is associated with a function $A_{W}(q,p)$ in phase space. This is achieved by using a linear map $\Omega_{W}: A\longrightarrow A_{W}(q,p)$, such that the associative algebra and the non-commutative structure of operators in $\mathcal{H}_E$ are preserved  in phase space. The consistence is due to the Weyl (or star) product of functions~\cite{Weyl}. Nevertheless, as the Wigner function is Real, the analysis of gauge symmetries  is compromised. This is an impediment to introducing the notion of interaction in the Wigner formalism (for the non-Abelian gauge symmetries, such as strong and weak interactions, the situation is even more complicated). The solution to this problem has been addressed by introducing a wave function, which is called a quasi-amplitude of probability, $\phi(q,p)$, from symplectic representations of symmetry groups.    In the case of Galilei  group, this led to the   symplectic quantum mechanics~\cite{Oliveira}.

In this  non-relativistic context,   $\phi(q,p)$ is a solution of the Schr\"{o}dinger equation represented phase space. The association with the Wigner function is given by $f_{W}(q,p)=\phi(q,p)\star\phi(q,p)^{\dag}$, where $\star$ is the Weyl product. As a representation theory, the symplectic quantum mechanics was generalized to analyze gauge symmetries, and for relativistics symmetries, it led to the Klein-Gordon and Dirac equations in phase space~\cite{Ronni2007}.

Exploring this symplectic theory, in the present work,  the Ginzburg-Landau free-energy is represented in a symplectic manifold with the phase space content. This is achieved by using the  concept of thermo group, the Lie-group structure associated with the thermofield dynamics and c*-algebras~\cite{PR2014,Santana2024}. Then a   symplectic unitary representation of the Euclidian group is studied. The order parameter is described by complex wave function $\phi(q,p)$. For abelian gauge symmetries, the self interaction and the interaction of the order parameter with an external field are addressed. Considering a superconductor,  well-known results are derived in order to show the consistency of the formalism. Beyond that, the structure of phase space provides other elements (not present in the usual formalism), such as the non-classical nature of the order parameter near to the critical temperature.

The presentation is organized as it follows. In Section \ref{sec2}, the notion of thermo-algebra is briefly reviewed and the notation is fixed. In Section \ref{sec3}, the thermo-Lie algebra for the Euclidian group is addressed.  In Section \ref{sec4}, the symplectic representation for the Euclidian symmetry leads to the Ginzburg-Landau theory formulated in phase space. The problem of abelian gauge symmetry is considered in  Section \ref{sec5}. In Section \ref{sec6}, the symplectic Ginzburga-Landau equations are presented and in Section \ref{sec7} aspects of symplectic current density are studied. Finally, in Section \ref{sec8} we discuss the quasi-probability distribution function with important indications of the non-classical nature of the system.  The concluding remarks are presented in Section \ref{sec9}.

% # new section ###############################
\section{Thermo algebra: definition and notation}\label{sec2}

One important aspect of thermal theories is a doubling in its algebraic structure. This is evident in thermofield dynamics, a real-time algebraic formalism for quantum field theories at finite temperature~\cite{Ademir2009}. In this case, the double Hilbert space is defined as  ${\cal H}_T={\mathcal H}_E\otimes \tilde{\cal H}_E$, where ${\mathcal H}_E$ and $\tilde{\mathcal H}_E$ are the  original and dual Hilbert space, respectively, both defined from the Euclidian manifold. Although not evident, this procedure of duplication is already present in the  Liouville-von Neumann equation usually given by
\begin{equation}
i\partial_{t}\rho(t)=\widehat{H}\rho(t),\nonumber
\end{equation}
where  $\rho(t)$ is the density matrix. For a pure state, $\rho$ is written as $\rho\simeq|\psi\rangle\langle \psi|$ and, for  a conservative system, the time evolution is given by the Liouvillian operator $\widehat{H}=[H,\cdot]$, where $H$ is the Hamiltonian of the system. The operator $\widehat{H}$ stands for the generator of time evolution, whilst $H$ is the observable describing energy of the system. This doubled structure is also present in kinetic theory. In algebraic terms, it was identified as the standard representations of the  c*-algebra. In the context of Lie-algebra it was called thermo-algebra~\cite{PR2014}. For the goal here, it is important to review some elements of this structure, in particular to fix the notation.

Let $L_{\mathcal{G}}=\{a_{i}, i=1,...,s\}$ be a Lie algebra, where  $a_{i}$ are the  symmetry generators of  the corresponding Lie group. The Lie product $(.,.)$ is written as
\begin{equation}\label{lie}
(a_{i},a_{j})=iC_{ij}^{k}a_{k},
\end{equation}
where a sum over repeated indices is assumed. The numbers $C_{ij}^{k}$ are the structure constants. The Lie product given in Eq.~(\ref{lie}) satisfies the anti-symmetric condition
\begin{equation}
(a_{i},a_{j})=-(a_{j},a_{i}),\nonumber
\end{equation}
and the Jacobi  identity,
\begin{equation}
(a_{i},(a_{j},a_{k}))+(a_{k},(a_{i},a_{j}))+(a_{j},(a_{k},a_{i}))=0.\nonumber
\end{equation}
The notion of thermo-Lie algebra is introduced as a representation of $L_{\mathcal{G}}$ for physical systems considering the following aspects~\cite{PR2014}. The set of kinematic variables, say $\mathcal{V}$, is a vector space defined  on a Hilbert space  $\mathcal{H}_{T}$, the carrier space of representation for the Lie-algebra $L_{\mathcal{G}}$. The set $\mathcal{V}$ is composed of two subspaces, that is, $\mathcal{V}=\mathcal{V}_{obs}\otimes\mathcal{V}_{gen}$. Here $\mathcal{V}_{obs}$ stands for the set of observables in a physical system and $\mathcal{V}_{gen}$ describes the symmetry generators. Considering  general aspects of physical symmetry~\cite{Ademir2009}, the representation of $L_{\mathcal{G}}$ in $\mathcal{H}_{T}$ is given by
\begin{eqnarray}\label{eqBT35}
[\widehat{A}_{i},\widehat{A}_{j}]&=&iC_{ij}^{k}\widehat{A}_{k},\nonumber\\
\ [\widehat{A}_{i},A_{j}]&=&iC_{ij}^{k}A_{k},\nonumber\\
\ [A_{i},A_{j}]&=&iC_{ij}^{k}A_{k},\nonumber
\end{eqnarray}
where $\widehat{A}\in \mathcal{V}_{gen}$ and $A \in \mathcal{V}_{obs}$ and $[\, ,\, ]$ is the commutator. This has been named a thermo-Lie algebra, $L^T_{\mathcal{G}}$,~\cite{PR2014}. In this context, it is important to observe that $L^T_{\mathcal{G}}$ is a semi-simple Lie-algebra. Then it is written as a direct sum of two Lie-algebras. Indeed, introducing the variable $\widetilde{A}=A-\widehat{A}$,   the Lie algebra $L^T_{\mathcal{G}}$  leads to
\begin{eqnarray}
[A_{i},A_{j}]&=&iC_{ij}^{k}A_{k}\label{A},\\
 \ [\widetilde{A}_{i},\widetilde{A}_{j}]&=&-iC_{ij}^{k}\widetilde{A}_{k}, \label{Atil2}\\
 \ [A_{i},\widetilde{A}_{j}]&=&0 \label{Atil233}.
\end{eqnarray}
This duplication is thought of as a mapping in $\mathcal{V}$, say $\mathcal{J}$: $\mathcal {V}\longrightarrow\mathcal{V}$, denoted by $JAJ=\widetilde{A}$. With this notation, and using Eqs.~(\ref{A})-(\ref{Atil233}),  the mapping $\mathcal{J}$ fulfills the conditions:
\begin{eqnarray}
(A_{i}A_{j})\widetilde{\ }&=&\widetilde{A}_{i}\widetilde{A}_{j},\nonumber\\
 \ (cA_{i}+A_{j}){\widetilde{\ }}&=&c^{*}\widetilde{A}_{i}+\widetilde{A}_{j},\nonumber\\
 \ (A_{i}^{\dag})\widetilde{\ }&=&(\widetilde{A}_{i})^{\dag},\nonumber\\
 \ (\widetilde{A}_{i})\widetilde{\ }&=&A_{i},\nonumber\\
 \ [A_{i},\widetilde{A}_{j}]&=&0.\nonumber
\end{eqnarray}
These properties are called   tilde conjugation rules in thermofield dynamics. In c*-algebra,  $\mathcal{J}$ is the modular conjugation~\cite{PR2014}.  In the next section, this representation is explored by considering the Euclidian group.

 % # new section ###############################
\section{Thermo Euclidian Lie-algebra }\label{sec3}

The Euclidian group is the semi-direct product of $O(3)\times T_{a}$ in $\mathbb{R}^{3}$, a subgroup of the Galilei group, containing the rotations, $O(3)$, and translations, $T_{a}$, preserving the norm. The associated Lie algebra is written as follows
\begin{eqnarray} \label{eq01}
  (l_{i},l_{j})&=&i\hbar\epsilon_{ijk}l_{k},\nonumber\\
\ (l_{i},p_{j})&=&i\hbar\epsilon_{ijk}p_{k},\nonumber
\end{eqnarray}
where $l_i$ is a generator of rotations and $p_i$ stand for the generator of space translations. The invariants are: $I_{1}=p^{2}$ and $I_{2}=l^{2}+s^{2}$, where $s$ stands for rotations in internal degrees of freedom in the carrier space. For this study we will take s = 0. The thermo algebra is written in terms of the non-zero commutation relationships among  observables and   tilde operators,
\begin{eqnarray}
[L_{i},L_{j}]&=&i\hbar\epsilon_{ijk}L_{k},\label{relcomut1}\\
\ [L_{i},P_{j}]&=&i\hbar\epsilon_{ijk}P_{k},\label{relcomut2}\\
\ [\tilde{L}_{i},\tilde{L}_{j}]&=&-i\hbar\epsilon_{ijk}\tilde{L}_{k},\label{relcomut3}\\
\ [\tilde{L}_{i},\tilde{P}_{j}]&=&-i\hbar\epsilon_{ijk}P_{k}\label{relcomut4},
\end{eqnarray}
highlighting the invariants $P^{2}$ and $\tilde{P}^{2}$.
The carrier space is $\mathcal{H}_{T}=\mathcal{H}_E\otimes\tilde{\mathcal{H}}_E$.  A vector in this thermal Hilbert space is denoted by $|\phi\rangle\in \mathcal{H}_{T}$.  The average value of an observable $\mathcal{A}$ with respect to $|\phi\rangle$ is written as $\langle\mathcal{A}\rangle=\langle \phi|\mathcal{A}|\phi\rangle$, with $\langle\phi|\phi\rangle=1$. A physical representation is constructed by introducing the vector $|I\rangle=\sum_{n}|n,\tilde{n}\rangle \in \mathcal{H} _{T}$. The state $|\phi\rangle \in \mathcal{H} _{T}$ is written in terms of an operator $F_{\phi} defined in \mathcal{H}_E$ as $ |\phi\rangle = (F_{\phi}\otimes1)|I\rangle=|F_{\phi}\rangle$. If $|\phi\rangle$ is normalized, then it leads to $Tr(F_{\phi}^{\dag}F_{\phi})=1$. At this point, it is convenient to introduce another notation by writing $F_{\phi}=\rho^{\frac{1}{2}}$. In other words,  we have
\begin{equation}
|\phi\rangle=|\rho^{\frac{1}{2}}\rangle=(\rho^{\frac{1}{2}}\otimes1)|I\rangle,\nonumber
\end{equation}
with $\rho^{\frac{1}{2}} \in \mathcal{H}_E$.

Now we are in a position to examine the action of the invariant   $\hat{P}^{2}=P^{2}-\tilde{P}^{2}$, on the vector $|\phi\rangle=|\rho^{\frac{1}{2}}\rangle$, where $\hat{P}$ is the space-translation generator. Since the tilde conjugation rule is an anti-linear mapping~\cite{Ademir2009}, $\hat{P}^{2}$ is written as  $\hat{P}^{2}=\nabla^{ 2}\otimes1-1\otimes\nabla^{2} = 0$. This leads to
\begin{eqnarray}
\hat{P}^{2}|\phi\rangle&=&\hat{P}^{2}(\rho^{\frac{1}{2}}\otimes 1  |I\rangle),\nonumber\\
&=&(\nabla^{2}\otimes 1 -1\otimes\nabla^{2})(\rho^{\frac{1}{2}}\otimes1|I\rangle), \nonumber\\
&=&[\nabla^{2}\rho^{\frac{1}{2}}-\rho^{\frac{1}{2}}\nabla^{2}]\otimes1|I\rangle,\nonumber\\
&=&[\nabla^{2},\rho^{\frac{1}{2}}]\otimes1|I\rangle,\nonumber\\
&=& 0.\nonumber
\end{eqnarray}
Hence, the operator $\rho^{\frac{1}{2}}(\phi)$ satisfies the following equation
\begin{equation}\label{LvN}
[\nabla^{2},\rho^{\frac{1}{2}}(\phi)]=0.
\end{equation}
Introducing the normalized operator $\rho=\rho^{\frac{1}{2}\dag}\rho^{\frac{1}{2}} $ as the density operator, we observe that
\begin{equation}\label{LvN2}
[\nabla^{2},\rho]=0.
\end{equation}
 This is a version of an time-independent Liouville-Von Neumann equation. In this case we are able to explore the phase-space Wigner representation. For a given operator $A$ defined in the Hilbert space $\mathcal{H}_E$, the linear Wigner mapping $\Omega_{W}:A\rightarrow A_{W}(q,p)$ is defined by~\cite{Oliveira}
\begin{equation}
     A_{W}(q,p)=\frac{1}{\sqrt{2\pi}}\int e^{-ipz}\left\langle q-\frac{z}{2}|A|q+\frac {z}{2}\right\rangle.\nonumber
\end{equation}
The product of two operators $AB$ in $\mathcal{H}_E$ is represented in phase space as
\begin{eqnarray}\label{eq15a4}
     (AB)_{W}(q,p)&=&A_{W}(q,p)e^{\frac{i\Lambda_0}{2}}B_{W}(q,p)\nonumber\\
      &=&A_{W} (q,p)\star B_{W}(q,p),
\end{eqnarray}
where $\Lambda_0=\overleftarrow{\partial_{q}}\overrightarrow{\partial_{p}}-\overleftarrow{\partial_{p}}\overrightarrow{\partial_{q}}$. The star-product defined in Eq.~(\ref{eq15a4})  is also called the Moyal product.  Here, the Planck constant is taken as $\hbar = 1$. The Wigner representation of Eq.~(\ref{LvN}) is derived by using the Wigner mapping, $\Omega_{W}$, i.e.,
\begin{eqnarray}\label{Mbracket1}
\Omega_{W}[\nabla^{2},\rho^{\frac{1}{2}}]&=&
\Omega_{W}[\nabla^{2}\rho^{\frac{1}{2}}]-\Omega_{W}[\rho^{\frac{1}{2}}
\nabla^{2}] ,\nonumber\\
&=&\ p^{2}\star f^{\frac{1}{2}}_{W}(q,p)-f^{\frac{1}{2}}_{W}(q,p)\star p^{2},\nonumber\\
&=&\{p^{2},f^{\frac{1}{2}}_{W}(q,p)\}_{M},\nonumber\\
&=&0,
\end{eqnarray}
where $\{a,b\}_{M}=a\star b-b\star a$ is the called Moyal bracket, with  $(q,p)$ are the coordinates of a symplectic manifold with phase space content and $f_W=\Omega_W(\rho)$. Proceeding  analogously with Eq.~(\ref{LvN2}), it leads to the Liouville-von Neumann equation in this symplectic manifold, namely
\begin{equation}\label{Mbracket2}
\{p^{2},f_{W}(q,p)\}_{M}=0.
\end{equation}
From Eqs.~(\ref{Mbracket1})~and~(\ref{Mbracket2}),  we observe that  $ f_{W(q,p)} \in \mathbb{R}$, that is, although normalized it can be negative. Indeed, observe that,
\begin{eqnarray}
f_W(q,p)&=&\Omega_{W}[\rho],\nonumber\\
&=&\Omega_{W}[\rho^{\frac{1}{2}\dag}\rho^{\frac{1}{2}}],\nonumber\\
f_{W}(q,p)&=&\psi(q,p)\star\psi^{\dag}(q,p),\nonumber
\end{eqnarray}
where we have used the notation $\Omega_{W}[\rho^{\frac{1}{2}}]=\rho^{\frac{1}{2}}_W(q,p)= \psi(q,p)$.

So far we have deduced a Wigner-like function on a symplectic manifold from the invariants of the thermo-Euclidian group. Thus, we will obtain an equation for the function $\psi(q,p)$.  This equation can be rederived by using a simplectic representation taking the cotangent structure associated with the Euclidian manifold. The resulting physical interpretation will be that $\psi(q,p)$ is the Ginzburg-Landau order parameter, but represented in phase space.

%# new section #####################################
\section{the symplectic Ginzburg-Landau order parameter}\label{sec4}

A symplectic structure associated to the Euclidian space, $\mathbb{E}$, is introduced through the notion of cotangent-bundle, $T^{*}\mathbb{E}$. We denote the coordinates of the Euclidian manifold, $\mathbb{E}$, by $q^{k}$, $k = 1, 2, 3$, and fix the metric $g=I$, where $I$ being the identity matrix in three dimensions. The coordinates of  $T^{*}\mathbb{E}$  are $(q^{k},p^{k})$, where $p^{k}$ is a coordinate in the dual, $\mathbb{E}^{*}$. The manifold $T^{*}\mathbb{E}$ is endowed with a symplectic structure by introducing  the 2-form $\omega= dq^{k}\wedge dp_{k}$ (a sum is assumed in the repeated indices).

We consider the space of analytical  functions  defined on the cotangent-bundle, $C^{\infty}(T^{*}\mathbb{E})$, and a linear operator defined by  $\Lambda_0=\overleftarrow{\partial_{q^{k}}}\overrightarrow{\partial_{p^{k}}}-\overleftarrow{\partial_{p^{k}}}\overrightarrow{\partial_{q_{k}}}$. For  $C^{\infty}$ functions $f(q,p)$ and $g(q,p)$, it is possible to write $\omega(f\Lambda,g\Lambda)=f\Lambda g=\{f,g\}$, where
\begin{equation}
\{f,g\}=\partial_{q^{k}}f\partial_{p_{k}}g-\partial_{p^{k}}f\partial_{q_{k}}g,\nonumber
\end{equation}
is the Poisson bracket. Equipped with this symplectic structure,  $T^{*}\mathbb{E}$ is the  phase space structure introduced in the previous section.  It will be denoted by $\Gamma_{E}$. Then a Hilbert space can be now introduced. In view of the linear subspace of integrable quadratic functions $\mathcal{H}(\Gamma)\subset C^{\infty}(\Gamma_{E})$, with complex values, $\{\phi: \Gamma_{E}\rightarrow \mathbb{C}\}$, that is,
\begin{equation}
    \int_{\Gamma_{E}}\phi^{\dag}(q,p)\phi(q,p)dqdp<\infty\nonumber
\end{equation}
where $q=(q^{1}, q^{2}, q^{3})$ and $p=(p^{1}, p^{2}, p^{3})$. Such a space becomes a Hilbert space by introducing the inner product
\begin{equation}\label{innerproduct}
    \langle\phi_{1}|\phi_{2}\rangle=\int_{\Gamma_{E}}\phi^{\dag}_{1}(q,p)\phi_{2}(q,p)dqdp.
\end{equation}
Furthermore, we consider the linear mappings $\bar{Q}$, $\bar{P}$: $\mathcal{H}(\Gamma)\rightarrow\mathcal{H}(\Gamma)$, such that $\bar{Q}\phi(q,p)=q\phi(q,p)$ and $\bar{P}\phi(q,p)=p\phi(q,p)$ . Since $[\bar{Q},\bar{P}]=0$, the spectrum of $\bar{Q}$ and $\bar{P}$ is used to construct a basis on $ \mathcal{H}(\Gamma),\{|q,p\rangle\}$, such that
\begin{equation}
  \bar{Q}|q,p\rangle=q|q,p\rangle,\ \ \ \bar{P}|q,p\rangle=p|q,p\rangle.\nonumber
\end{equation}
Since $\phi(q,p)=\langle q,p|\phi\rangle$, we derive from Eq.~(\ref{innerproduct})  the result
\begin{equation}
    \int |q,p\rangle\langle q,p|dqdp=1\nonumber
\end{equation}
with $\langle q,p|q',p'\rangle=\delta(q-q')\delta(p-p')$. It is worth emphasizing  that the operators $\bar{Q}$ and $\bar{P}$ are not the usual quantum mechanical operators for position and momentum, since we are dealing with a symplectic structure for the field $\phi(q,p)$, which will be considered an order parameter.

Using the $q$ and $p$ functions defined in $\mathcal{H}(\Gamma)$, we build the following tilde and non-tilde operators, i.e.,
\begin{eqnarray}
P=p{\star}&=&p-\frac{i\hbar}{2}\partial_{q},\nonumber\\
Q=q{\star}&=&q+\frac{i\hbar}{2}\partial_{p},\nonumber\\
\widetilde{P}={\star}p&=&p+\frac{i\hbar}{2}\partial_{q},\nonumber\\
\widetilde{Q}={\star}q&=&q-\frac{i\hbar}{2}\partial_{p},\nonumber
\end{eqnarray}
and also
\begin{eqnarray}
L_{i}&=&\epsilon_{ijk}Q_{j}P_{k},\nonumber\\
\widetilde{L}_{i}&=&\epsilon_{ijk}\widetilde{Q}_{j}\widetilde{P}_{k}.\nonumber
\end{eqnarray}
This set of operators satisfies the same commutation relations as presented in Eqs.~(\ref{relcomut1})-(\ref{relcomut4}). As a consequence, $Q$ and $P$ are transformed as position and momentum, and, are by construction observables due to the structure of the thermo algebra. Taking the invariant $P^{2}=(p\star)^{2}=\alpha$, it follows 
\begin{equation}\label{GLeq}
\frac{1}{m}\left(p^{2}-i\hbar p\partial_{q}-\frac{\hbar^{2}}{4}\partial_{q}^{2}\right)\phi(q,p)=\alpha\phi(q,p).
\end{equation}
This equation is the Ginzburg-Landau equation without the nonlinear term written in a symplectic manifold through the representation of the Euclidian group. This is valid since we consider $\alpha=\alpha_{0}(T-T_{c})=\alpha(T)$, where $T_{c}$ is the critical temperature of the phase transition. The field $\phi(q,p)$ is the order parameter with phase space content, which is associated with a quasi-distribution (Wigner-like) function given by  that is
\begin{equation}
f(q,p)=\phi(q,p)\star\phi^{\dag}(q,p).\nonumber
\end{equation}
The average of an observable $A$ is given by
\begin{eqnarray}
\langle A\rangle&=&\int f(q,p)A_W(q,p)dqdp,\nonumber\\
&=&\int \phi^{\dag}(q,p)A(Q,P)\phi(q,p)dqdp.\nonumber
\end{eqnarray}
Equation~(\ref{GLeq}) is obtained from the action
\begin{eqnarray}
S&=&\int\mathcal{L}_{0}dpdq,\nonumber\\
&=&\int\left[\frac{1}{m}(p\star\phi)^{\dag}(p\star\phi)+\alpha(T)\phi^{\dag}\phi\right]dpdq,\nonumber
\end{eqnarray}
where the free energy density, $\mathcal{L}_{0}$,  is given by
\begin{equation}\label{Lagrangina0}
\mathcal{L}_{0}=\frac{1}{m}(p\star\phi)^{\dag}(p\star\phi)+\alpha(T)\phi^{\dag}\phi.
\end{equation}
For a self-interaction, a term preserving the invariance of $\mathcal{L}$ by $\phi\rightarrow -\phi$, is introduced by the phase space density, $\phi^{\dag}(q,p)\phi(q,p)$, like as $\frac{\beta}{2}(\phi^{\dag}\phi)^2$. The free-energy density is now written as
\begin{equation}
\mathcal{L}=\frac{1}{m}(p\star\phi)^{\dag}(p\star\phi)+\alpha(T)(\phi^{\dag}\phi)+\frac{\beta}{2}(\phi^{\dag}\phi)^2.\nonumber
\end{equation}
The Euler-Lagrange equation for $\phi^{\dag}(q,p)$ is
\begin{equation}
\frac{\partial\mathcal{L}}{\partial \phi^{\dag}}-\partial_{q}\left[\frac{\partial\mathcal{L}}{\partial(\partial_{q}\phi^{\dag})}\right]-\partial_{p}\left[\frac{\partial\mathcal{L}}{\partial(\partial_{p}\phi^{\dag})}\right]=0.\nonumber
\end{equation}
This leads to the following equation for $\phi(q,p)$,
\begin{equation}\label{EqGlA=0}
\frac{1}{m}\left(p-\frac{i\hbar}{2}\partial_{q}\right)^{2}\phi+\alpha(T)\phi+\beta|\phi|^{2}\phi=0.
\end{equation}
In order to find a solution, we assume the ansatz $\phi(q,p)=\exp\left(-\frac{2i}{\hbar}qp\right)F(p)\varphi(q)$ to write
\begin{equation}
-\frac{\hbar^{2}}{2m^{*}}\frac{d^{2}}{dq^{2}}\varphi(q)+\alpha_{0}(T-T_{c})\varphi(q)+\beta_{p}\varphi^{3}(q)=0,\nonumber
\end{equation}
where $m^{*}=2m$, $\alpha(T)=\alpha_{0}(T-T_{c})$ and $\beta_{p}=\beta|F(p)|^{2}$. Following the usual procedure and for the superconducting regime ($T<T_{c}$), we define the coherence length $\xi^{2}\equiv\frac{\hbar^{2}}{2m^{*}|\alpha|}$, arriving at the solution
\begin{equation}\label{solu_GL}
\varphi(q)=\sqrt{\frac{|\alpha(T)|}{\beta_{p}}}\tanh\left[\frac{q}{\xi(T)\sqrt{2}}\right].
\end{equation}
The complete solution given by
\begin{eqnarray}\label{solu_particular}
\phi(q,p)&=&\exp\left(-\frac{2i}{\hbar}qp\right)\frac{F(p)}{|F(p)|}\nonumber\\
&&\times\sqrt{\frac{|\alpha(T)|}{\beta}}\tanh\left[\frac{q}{\xi(T)\sqrt{2}}\right].
\end{eqnarray}
The density is written as 
\begin{equation}\label{solu_GL2}
|\phi(q,p)|^{2}=|\varphi(q)|^{2}=\frac{|\alpha(T)|}{\beta}\tanh^{2}\left[\frac{q}{\xi(T)\sqrt{2}}\right].
\end{equation}
Hence, this symplectic representation reproduces the  result of the Ginzburg-Landau theory for the density of the order parameter~\cite{Ketterson}. 

In the following the abelian gauge symmetry is taken into consideration.

 %# new section #####################
\section{Abelian Gauge Symmetry}\label{sec5}

The abelian gauge associated with the symplectic Ginzburg-Landau order parameter is introduced following standard procedures~\cite{1Ronni2018,2Ronni2018,Ronni2019}. Considering initially the free theory, the Lagrangian density $\mathcal{L}_{0}$ given in  Eq.~(\ref{Lagrangina0}) is analyzed under local gauge transformations,  $\phi(q,p) \rightarrow e^{-i\Lambda(q,p)}\star \phi(q,p)$.

For $\Lambda\ll 1$, it leads to
\begin{eqnarray}
\delta\phi&=&-i\Lambda\star\phi,\nonumber\\
\delta\phi^{\dag}&=&i\phi^{\dag}\star\Lambda.\nonumber
\end{eqnarray}
The Lagrangian density variation on the local gauge is given by
\begin{equation}
\delta\mathcal{L}_{0}=\phi^{\dag}\star p^{k}\left(\partial_{q^{k}}\Lambda\right)\star\phi.\nonumber
\end{equation}
Then the free-energy density in Eq.~(\ref{Lagrangina0}) is not invariant by local gauge transformations. The invariance is achieved by introducing a new field $A^{k}$, the gauge field, through a term $A^{k}\star\phi(q,p)$. The field $A^{k}(q,p)$ is required to transform as
\begin{equation}
A^{k}\rightarrow A'^{k}=A^{k}+i\{A^{k},\Lambda\}_{M}-\frac{c\hbar}{e}\left(\partial_{q^{k}}\Lambda\right),\nonumber
\end{equation}
where $e$ is the field coupling constant and $\{a,b\}_{M}=a\star b - b\star a$ is the aforementioned Moyal bracket. Thus the star operator $p^{k}_{\star}$ is replaced by the covariant derivative $D^{k}_{\star}$
\begin{equation}
D^{k}\star=p^{k}\star-\frac{e}{c}A^{k}\star,\nonumber
\end{equation}
and the corresponding adjoint operator,
\begin{equation}
\star D^{k}=\star p^{k}-\frac{e}{c}\left(\star A^{k}\right),\nonumber
\end{equation}
where $\phi^{\dag}\star p=\left(p+\frac{i\hbar}{2}\partial_{q}\right)\phi^{\dag}$. This leads to a free-energy that is gauge-invariant at the critical point, $\alpha(T)\rightarrow\alpha(T_c)=0$, i.e.,    
\begin{equation}
\mathcal{L}=\frac{1}{m}\left[(D^{k}\star\phi)^{\dag}(D^{k}\star\phi)\right].\nonumber
\end{equation}
 The gauge field $A^{k}$ provides an extra term to the free-energy density, i.e.,  $F^{kl}F_{kl}$, where $F^{kl}$ is the field tensor represented in phase space and given by
\begin{equation}
F^{kl}=\partial_{q^{k}}A^{l}-\partial_{q^{l}}A^{k}+\frac{ic\hbar}{e}\{A^{l},A^{k}\}_{M}.\nonumber
\end{equation}
The  full free-energy density is written as
\begin{equation}\label{Lagragiana1}
\mathcal{L}=\frac{1}{m}\left[(D^{k}\star\phi)^{\dag}(D^{k}\star\phi)\right]-\frac{1}{c}F^{kl}F_{kl},
\end{equation}
which is gauge-invariant at the critical point. This result is a realization of the non-commutative gauge field theory, which was proposed by  Seiberg-Witten~\cite{SeibergWitten1999,Ronni2015}. In our case, $A^{k}$ is a phase space representation of the magnetic potential. This fact was proved in general for the  relativistic symplectic representation of spin 1.~\cite{Ronni2015} This result  will be evident in the calculation of the current density.

 %# new section ########################
\section{Ginzburg-Landau symplectic equations}\label{sec6}

Starting from the Lagrangian density, invariant by local gauge transformations, Eq.~(\ref{Lagragiana1}), plus the Ginzburg-Landau invariant terms, we write the Symplectic Ginzburg-Landau Lagrangian density,
\begin{eqnarray}\label{den-lagrangiana-mod-acres}
\mathcal{L}_{SGL}&=&\frac{1}{m}(D^{k}\star\phi)^{\dag}(D^{k}\star\phi)-\frac{1}{c}F^{kl}F_{kl}+\alpha(T)\phi^{\dag}\phi\nonumber\\
&&+\frac{\beta}{2}(\phi^{\dag}\phi)^{2}.
\end{eqnarray}
Therefore, two symplectic equations are derived from the Euler-Lagrange equations for the fields $\phi^{\dag}(q,p)$, i.e.
\begin{equation}\label{EqSGL1a}
\frac{1}{m}\left(\star D\right)^{2}\phi^{\dag}(q,p)+\alpha(T)\phi^{\dag}+\beta|\phi(q,p)|^{2}\phi^{\dag}(q,p)=0,
\end{equation}
and for $\phi(q,p)$,
\begin{equation}\label{EqSGL1b}
\frac{1}{m}\left(D\star\right)^{2}\phi(q,p)+\alpha(T)\phi+\beta|\phi(q,p)|^{2}\phi(q,p)=0.
\end{equation}
The latter is reduced to Eq.~(\ref{EqGlA=0}), by taking $A=0$, in the star operator $D\star=p\star-\frac{e}{c}A\star $, which was resolved to a particular solution. Equations~(\ref{EqSGL1a}) and (\ref{EqSGL1b}) are, therefore, the first symplectic Ginzburg-Landau equations.

The  Euler-Lagrange equation for the gauge field $A^{k}$ is given by
\begin{equation*}
\frac{\partial\mathcal{L}_{GLS}}{\partial A^{k}}-\partial_{q^{l}}\left[\frac{\partial\mathcal{L}_{GLS}}{\partial \left(\partial_{q^{l}}A^{k}\right)}\right]-\partial_{p^{l}}\left[\frac{\partial\mathcal{L}_{GLS}}{\partial \left(\partial_{p^{l}}A^{k}\right)}\right]=0.
\end{equation*}
It leads to
\begin{eqnarray*}
\partial_{q^{l}}F^{kl}(q,p)={\cal J}^{k}(q,p),
\end{eqnarray*}
with ${\cal J}^{k}(q,p)$ given by
\begin{equation}\label{J}
{\cal J}^{k}(q,p)=\frac{e}{m}\left[\left(\phi^{\dag}\star D^{k}\right)\phi+\phi^{\dag}\left(D^{k}\star\phi\right)\right].
\end{equation}
Such an expression for the current density  ${\cal J}^{k}(q,p)$ is configured as the second symplectic Ginzburg-Landau equation. This current density in phase space has zero divergence, so it is enough to verify that, since the tensor $F^{kl}$ is anti-symmetric, it immediately implies $\partial_{q^{k}}\partial_{q^{l} }F^{kl}=\partial_{q^{k}}{\cal J}^{k}=0$. The zero divergence of the deduced current density is obtained directly by deriving the coordinate $q$, and by using Eqs.~(\ref{EqSGL1a}) and (\ref{EqSGL1b}), for the fields $\phi$ and $\phi^{\dag}$. Considering $A\star\phi=A\phi$ and $\phi^{\dag}\star A=\phi^{\dag}A$,  the symplectic current density is rewritten as
\begin{equation*}
{\cal J}^{k}(q,p)=\frac{e}{m}\left[\left(\phi^{\dag}\star p^{k}\right)\phi+\phi^{\dag}\left(p^{k}\star\phi\right)-\frac{2e}{c}A^{k}|\phi|^{2}\right].
\end{equation*}
With the star operators $\star p$ and $p\star$, it leads to
\begin{eqnarray*}
{\cal J}^{k}(q,p)&=&-\frac{ie\hbar}{2m}\left[\phi^{\dag}\left(\partial_{q^{k}}\phi\right)-\left(\partial_{q^{k}}\phi^{\dag}\right)\phi\right]\\
&&-\frac{2e^{2}}{mc}\left(p^{k}-\frac{e}{c}A^{k}\right)|\phi|^{2}.
\end{eqnarray*}

Writing $\phi(q,p)=\exp\left(-\frac{2i}{\hbar}qp\right)F(p)\sigma(q)$ and performing the appropriate derivations, we obtain
\begin{equation}
{\cal J}(q,p)=|F(p)|^{2}\left\{-\frac{i\hbar e}{2m}\left[\sigma^{\dag}{\bf \nabla}\sigma-\sigma{\bf \nabla}\sigma^{\dag}\right]-\frac{2e^{2}}{mc}{\bf A}|\sigma|^{2}\right\}.\nonumber
\end{equation}
Integrating in $p$, considering $|F(p)|^{2}$ as a Dirac delta distribution $\delta(p)$, we obtain
\begin{equation}
 {\bf J}_{GL}(q)=-\frac{i\hbar e}{2m}\left(\sigma^{\dag}{\bf \nabla}\sigma-\sigma{\bf \nabla} \sigma^{\dag}\right)-\frac{2e^{2}}{mc}{\bf A}|\sigma|^{2},\nonumber
\end{equation}
which is the usual Ginzburg-Landau current density in the configuration (Euclidian) space.~cite{Ketterson,Schmidt} From this result for the usual preserved Ginzburg-Landau current density, it is possible to deduce the characteristic effect of magnetic shielding, known as the Meissner-Ochsenfeld effect, and also to determine the maximum external magnetic field supported by the superconductor.

In the next, we demonstrate the proportionality of the critical current density with temperature, directly from the symplectic Lagrangian density.

%# new section ########################
\section{Critical Current Density}\label{sec7}

From the Ginzburg-Landau symplectic Lagrangian density, Eq.~(\ref{den-lagrangiana-mod-acres}), without  any \textit{a priori} approximation, we have
\begin{eqnarray*}
\mathcal{L}_{SGL}&=&\left(p-\frac{e}{c}A\right)\frac{i\hbar}{2m}\left[\left(\partial_{q}\phi^{\dag}\right)\phi-\phi^{\dag}\left(\partial_{q}\phi\right)\right]\\
&&+\left[\alpha(T)+\frac{1}{m}\left(p-\frac{e}{c}A\right)^{2}\right]|\phi|^{2}+\frac{\beta}{2}\left(|\phi|^{2}\right)^{2},\nonumber
\end{eqnarray*}
where, again, we choose the simplest relation of the action of $A\star$, that is, $A\star\phi=A\phi$ and $\phi^{\dag}\star A=\phi^{\dag}A$. Deriving the Lagrangian density, relative to $|\phi|^{2}$, we look at   
$\frac{\partial\mathcal{L}_{SGL}}{\partial|\phi|^{2}}=0$, that leads to
\[
\alpha(T)+\frac{1}{m}\left(p-\frac{e}{c}A\right)^{2}+\beta|\phi|^{2}=0.
\]
Therefore, we obtain
\begin{eqnarray}
|\phi|^{2}=\frac{|\alpha(T)|}{\beta}\left[1-\frac{mv_{\phi}^{2}}{|\alpha(T)|}\right],\nonumber
\end{eqnarray}
where $v_{\phi}$ is the field flow velocity, given by $v_{\phi}=\frac{1}{m}\left(p-\frac{e}{c}A\right)$. In this case, the expression for the current density, Eq.~(\ref{J}), is rewritten as
\begin{equation}\label{Jaberto}
{\cal J}(q,p)=-\frac{ie\hbar}{2m}\left[\phi^{\dag}\left(\partial_{q}\phi\right)-\left(\partial_{q}\phi^{\dag}\right)\phi\right]+\frac{2e}{m}\left(p-\frac{e}{c}A\right)|\phi|^{2}.
\end{equation}

The variation of the field $\varphi(q)$ occurs in the direction perpendicular to the flow moment $p$, somewhat similar to the stationary laminar flow of a viscous fluid. In this sense, the squared modulus of the field $\varphi(q)$, which is interpreted in the usual theory as being the density of superconducting charge carriers, stands for the velocity profile of the field flow $\phi(q,p) $. The flow, however, is established in the perpendicular direction of the velocity profile and is represented by the coordinate $p$. Thus, in Eq.~(\ref{Jaberto}) above, we can neglect the variation of this profile when looking at the variation in flow. In general, the two variations are linked, but can be conveniently decoupled. In this way, we rewrite the symplectic current density as a function of the field flow velocity, that is,
\begin{eqnarray*}
{\cal J}(v_{\phi})&=&2ev_{\phi}|\phi|^{2},\\
&=&2e\frac{|\alpha(T)|}{\beta}\left[1-\frac{mv_{\phi}^{2}}{|\alpha(T)|}\right]v_{\phi}.
\end{eqnarray*}
When derived by concerning the same flow velocity of the field, we arrive at the following result pointed out by the usual phenomenological theory, i.e., $\frac{1}{2}m^{*}v_{\phi}^{2}=\frac{|\alpha (T)|}{3}$, where $m^{*}=2m$. Using this into the expression for symplectic current density, we obtain the critical current density, the maximum current supported by the superconductor, a result also found by the usual theory, namely
\begin{eqnarray*}
{\cal J}_{c}&=&\frac{4e}{3\beta\sqrt{3m}}|\alpha(T)|^{\frac{3}{2}}.
\end{eqnarray*}
In conclusion, we re-derive the proportionality with the temperature of the critical current density, $\propto |\alpha(T)|^{\frac{3}{2}}$, demonstrating once again the consistency of the symplectic representation. This result also points to a certain critical moment of the field $\phi(q,p)$, considering the absence of the gauge field $A$, namely
\begin{equation}\label{pc}
    p_{c}=\frac{\hbar}{2\sqrt{3}\xi(T)}.
\end{equation}

In the next section, the quasi-probability distribution function is analyzed for the particular solution of Section~\ref{sec4}.

% # new section ###############################
\section{The Quasi-distribution Function of Probability}\label{sec8}

In order to calculate the quasi-probability distribution function, we consider the particular solution $\phi(q,p)$ obtained in Section~\ref{sec4} for the first symplectic Ginzburg-Landau equation with $A=0$, Eq.~(\ref{solu_particular}). Furthermore, we start from the required form of $|F(p)|^{2}$ in Section~\ref{sec6}, taken as a Dirac delta distribution, $\delta(p)$, which is represented as a Gaussian, that is,
\begin{equation*}
|F(p)|^{2}=\lim_{\frac{\xi(T)}{\hbar}\rightarrow \infty}\frac{\xi(T)}{\hbar\sqrt{\pi}}\exp\left(-\frac{\xi^{2}(T)}{\hbar^{2}}p^{2}\right).
\end{equation*}
Hence, we also have 
\begin{equation*}
\frac{F(p)}{|F(p)|}=\exp\left(\frac{i\xi^{2}(T)}{\hbar^{2}}p^{2}\right) 
\end{equation*}
and the respective conjugated function
\begin{equation*}
\frac{F^{\dag}(p)}{|F(p)|}=\exp\left(-\frac{i\xi^{2}(T)}{\hbar^{2}}p^{2}\right).
\end{equation*}
% Figura 01 #################################################
\begin{figure}[b]
\centerline{\includegraphics[width=6.8cm]{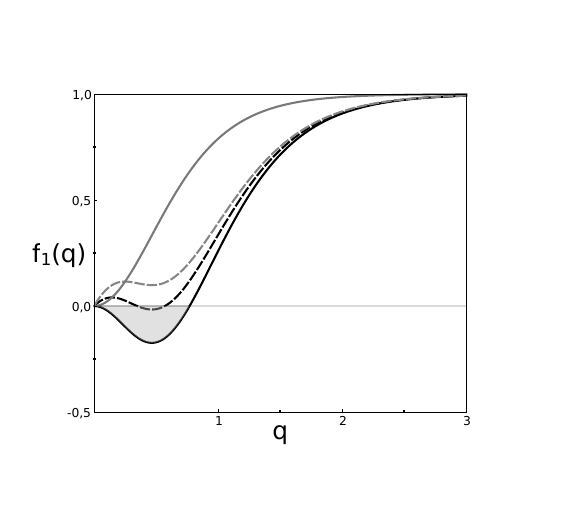}}
\caption{Plots of $f_1(q,p)$ as given in Eq.~(\ref{f1Wigner}) for different values of $p$. Gray dashed line: $|\varphi(q)|^{2}$ stands for the  result derived in the usual Ginzburg-Landau density. Black solid line: $f_{1}(q,p=0)$. Black dotted line: $f_{1}(q,p=p_{c})$. Black dashed line: $f_{1}(q,p=1)$. \label{f01}}
\end{figure}
%########################################################

From the quasi amplitude $\phi(q,p)$, the quasi-distribution function $f(q,p)$ is calculated  by expanding the star product, that is,  
\begin{equation}
f(q,p)= \phi(q,p)\left[1+\frac{i\hbar}{2}\Lambda_{0} +\frac{1}{2!}\left(\frac{i\hbar}{2}\Lambda_{0}\right)^{2}+\cdot\cdot\cdot\right]\phi^{\dag}(q,p).\nonumber
\end{equation}
Computing terms up to first order in $\hbar$, it leads to 
\begin{eqnarray*}
f_{1}(q,p)&=& \phi(q,p)\left(1+\frac{i\hbar}{2}\Lambda_{0}\right)\phi^{\dag}(q,p),\\
 &=&|\varphi(q)|^{2}-\left[q-\frac{\xi^{2}(T)}{\hbar}p\right]\frac{d}{dq}|\varphi(q)|^{2}.
\end{eqnarray*}
In order to analyze the behavior of such a solution, we take  $|\alpha(T)|=\beta=\hbar=m=1$,  leading to 
\begin{eqnarray}
f_{1}(q,p)=\tanh^{2}(q\sqrt{2})\left[1-\frac{4\sqrt{2}\left(q-\frac{p}{4}\right)}{\sinh\left(2q\sqrt{2}\right)}\right]. \label{f1Wigner}
\end{eqnarray}
The behavior of this quasi-distribution function is presented in Figure~\ref{f01}. We first compare   $f_{1}(q,p)$, for some fixed values of $p$ with  the standard Ginzburg-Landau density, $|\phi(q)|^{2}$. That is, the standard behavior for a typical superconductor is obtained at the zeroth order of $\hbar$ (gray solid line).  The deduced quasi-distribution function reaches a negative region close to the edge of the superconductor, even with $p=0$ (black solid line), where a non-classical behavior of the system is identified as a consequence of he negativity factor. Geometrically this factor is represented here as the negative gray-area indicated in Figure~\ref{f01}\cite{Santana2024}.    We note, however, that the negativity factor is maximum for $p=0$ and cancels out for 
\begin{equation*}
    p=\frac{\hbar}{\xi(t)}\sqrt{2}\left[\cosh^{-1}\left({\sqrt{\frac{3}{2}}}\right)-\frac{\sqrt{3}}{4}\right],
\end{equation*}
slightly above $p_{c}$, Eq.~(\ref{pc}), pointing to the loss of non-classicity. In addition, it is also possible to observe that the negative region accessed is closer to the edge of the material, indicating that the  system, in this region, presents non-classical behavior. 

 %# new section ###########################
\section{Final concluding remarks}\label{sec9}

In short, an Euclidean representation for the Ginzburg-Landau theory is developed by using a   Hilbert space defined in  a symplectic manifold. First, with the thermal algebra approach and the tilde conjugation rules, a  Liouville-Von-Neumann-like  equation is derived. Then, we identify the order parameter of the phenomenological theory with a quasi-amplitude of probability, which is associated with a  quasi-probability distribution function, i.e. a Wigner-type function. The first Ginzburg-Landau equation in phase space is derived in the absence of an external field. We advanced in building the formalism in a more robust way and wrote the gauge theory, from which we obtained the expression for the superconducting current density. We show that the quasi-distribution function, $f_{\phi}(q,p;T)$, associated with the solution $\phi(q,p)$, presents a behavior that diverges from the usual charge carrier density, with a profile reaching negative regions of the domain, closer to the edge of the superconducting material. These negative regions, which are indicating a non-classicality structure of the system, are present for certain values of the momentum associated with the field $\phi(q,p)$, canceling out for a specific value of the coordinate $p$. What was possible to infer from the results is that this limit value for the moment associated with the field is certainly very close to the value found that determines the unpairing of superconducting charge carriers; in the literature called critical moment.

 Each step of our approach emphasizes the concern with the construction of a theory of representation of symmetry groups. The association of a Wigner-type function with the order parameter of the Ginzburg-Landau theory allows us to establish a general parallel with the usual theory. The calculations, in general, identify the typical behavior of a superconductor, as its characteristic lengths, and its physical dynamics, characteristic of a phenomenological mean field theory.

The symplectic representation of the Ginzburg-Landau theory still presents sectors to be explored. We hope to determine new solutions for the symplectic equations and see what new information can be obtained. The non-abelian  representation is also to be developed. The representation of this formalism in a symplectic Fock space (corresponding to a quantization of the theory, equivalent to a many-body formalism) and the determination of correlation functions deserves  certainly to be explored. These aspects will be presented elsewhere.

\section*{Acknowledgments}
 This work partially  supported by CNPq, a Brazilian Federal Government Agency. The Authors thank to Renato Luz, Alesandro F. Santos and Rendislei Paiva for the discussions.
% Bibliografia####################

% Fim do documento ########################
\end{document}